\documentclass[lettersize,journal]{IEEEtran}
\usepackage{amsmath,amsfonts}
\usepackage{algorithmic}
\usepackage{algorithm}
\usepackage{array}
\usepackage[caption=false,font=normalsize,labelfont=sf,textfont=sf]{subfig}
\usepackage{textcomp}
\usepackage{stfloats}
\usepackage{url}
\usepackage{verbatim}
\usepackage{graphicx}
\usepackage{cite}
\usepackage{amsmath,amssymb,amsfonts}
\usepackage{algorithmic}
\usepackage{graphicx}
\usepackage{graphics}
\usepackage{afterpage}
\usepackage{textcomp}
\usepackage{xcolor}
\usepackage{hyperref}
\usepackage{float}
\usepackage{svg}

\begin{document}

\title{CGoT: A Novel Inference Mechanism for Embodied Multi-Agent Systems Using Composable Graphs of Thoughts}



\author{Yixiao Nie, Yang Zhang, Yingjie Jin, Zhepeng Wang, Xiu Li, and Xiang Li*
\thanks{Y. Nie, Y. Zhang and Xiu Li is with the Shenzhen Tsinghua International Graduate School, Tsinghua University. Y. Jin and Z. Wang are with Lenovo Research.
Xiang Li is with the Department of Automation, Tsinghua University. Project leader: Yang Zhang. Corresponding author: Xiang Li.}%
}

\maketitle

\begin{abstract}
The integration of self-driving cars and service robots is becoming increasingly prevalent across a wide array of fields, playing a crucial and expanding role in both industrial applications and everyday life. In parallel, the rapid advancements in Large Language Models (LLMs) have garnered substantial attention and interest within the research community. This paper introduces a novel vehicle-robot system that leverages the strengths of both autonomous vehicles and service robots. In our proposed system, two autonomous ego-vehicles transports service robots to locations within an office park, where they perform a series of tasks. The study explores the feasibility and potential benefits of incorporating LLMs into this system, with the aim of enhancing operational efficiency and maximizing the potential of the cooperative mechanisms between the vehicles and the robots. This paper proposes a novel inference mechanism which is called CGOT toward this type of system where an agent can carry another agent. Experimental results are presented to validate the performance of the proposed method.
\end{abstract}



\section{Introduction}

Autonomous ego-vehicles have been increasingly applied across various domains \cite{ni2020survey}. For instance, they serve as daily vehicles for families \cite{badue2021self}, while in public transportation, they are deployed as self-driving taxis and buses, contributing to reduced labor costs and enhanced safety \cite{woodward2021intelligent}. In industrial settings, ego-vehicles are employed to transport goods, thereby improving production efficiency \cite{chowdhury2020trustworthiness}. Concurrently, significant research has focused on advancing autonomous driving technologies to address the complexities of dynamic driving environments and road conditions. These efforts encompass critical areas such as visual perception \cite{b2,ni2020survey} and motion planning \cite{gonzalez2015review}, which are essential for the reliable operation of self-driving systems \cite{b1}.

Most current studies view autonomous vehicles as independent systems confined to transport functions. In contrast, this paper proposes a different perspective: the ego-vehicle acts as a mobile platform carrying service robots that extend its capabilities. Upon reaching a target location, the vehicle deploys the robots to perform tasks such as package delivery and retrieves them after completion. It then proceeds to the next site, repeating the cycle. This paradigm enhances the system’s versatility by enabling diverse missions, improves efficiency through parallel operations, and extends the effective working range by allowing robots to act beyond the vehicle’s immediate surroundings.

Simultaneously, Large Language Models (LLMs) have garnered significant attention since their rapid development began in recent years. Following the release of GPT-3.5, numerous LLMs have been introduced and integrated into various task scenarios, demonstrating their versatility across multiple domains. Among these applications, embodied agents stand out as a particularly promising area of research. Various approaches have been explored to enhance the performance of LLM-based systems, with recent studies indicating that while LLMs may not solve all planning-related challenges, appropriate modifications to system design can substantially improve their problem-solving capabilities.


In this work, we propose a delivery assistance system for a workplace park that incorporates multiple agents, each with distinct features and capabilities tailored to specific tasks in different areas of the park. For instance, while the ego-vehicle is designed for extended operational range and duration, the mobile robots are smaller and more agile, enabling them to perform complex maneuvers and access hard-to-reach locations. We have also integrated LLMs into the system to assist the agents in 
constructing graphs that represents the inferences of the agents during the task processing.

\section{Related Work}
\subsection{Ego-vehicles and Service Robots}
Several existing works have explored systems that facilitate cooperation between robots. For example, Lee et al. proposed a system where UAVs guide ground robots with limited visibility \cite{lee2023aerial}; Xu et al. focused on the docking systems for Autonomous Underwater Vehicles (AUVs) \cite{xu2024stereo}, where the docking station remains stationary.

The field of traditional robotic mobile control has matured significantly, with path planning algorithms like A* \cite{hart1968formal}, Dijkstra's \cite{dijkstra2022note}, and Minimum Spanning Tree \cite{prim1957shortest} playing crucial roles. Finite state machines have also been used to dictate robot actions in various states. Additionally, the fusion of data from sensors such as laser rangefinders, cameras, and inertial measurement units has improved robots' perception and localization capabilities \cite{yuan2019multisensor,teng2023fusionplanner}. While these conventional methods are effective in static environments, modern mobile robots in complex and dynamic settings often combine these techniques with advanced methods like machine learning and deep learning to enhance planning and control. Machine learning-based approaches in motion planning and control are categorized into three types: end-to-end learning systems that map perceptual inputs to motion commands \cite{bojarski2016end,codevilla2018end,zhang2016query}, methods that enhance specific subsystems like global or local planning \cite{xiao2021agile}, and techniques that improve components of traditional navigation systems to boost performance and robustness \cite{xiao2020appld,stein2018learning}. However, for self-driving systems operating in complex environments, such as the one proposed in this paper, existing learning-based methods may not always be the optimal solution.
\subsection{LLMs Application}

With the rapid advancement of Large Language Models (LLMs), an increasing number of tasks across various domains have begun to leverage these models to enhance performance and unlock untapped potential. Researchers like H. Zhang et al. and C. Zhang et al. have explored innovative approaches to integrating LLMs into multi-agent systems, combining them with diverse modules such as communication, memory, and perception to create more cohesive and effective systems \cite{zhang2023building, zhang2024proagent}. Lin et al. further investigated the potential of merging LLM-based planning with traditional methods, demonstrating how this hybrid approach can lead to more efficient and effective problem-solving strategies in complex planning scenarios \cite{lin2024swiftsage}.

In addition to planning, controlling tasks with the assistance of LLMs has emerged as a cutting-edge area of research. For instance, \cite{gong2023lemma} focused on learning language-conditioned multi-robot manipulation, while \cite{mandi2024roco} conducted studies on cooperative systems capable of performing zero-shot tasks through the implementation of LLMs. Moreover, Liu et al. \cite{liu2022embodied} explored the integration of scene-level visual perception with task plan generation, enabling the system to interpret ambiguous instructions and generate detailed task plans using LLMs. Huang et al. \cite{huang2022language} considered the application of LLMs as zero-shot human models, significantly enhancing human-robot interactions by providing robots with a better understanding of human behaviors and intentions.

Additionally, there has been considerable research focused on breaking down complex tasks into smaller, more manageable sub-tasks at the initial stages of the planning process. Notable works have contributed to this approach\cite{ahn2022can,song2023llm,besta2024graph}, while Yao et al. studied ways to boost system efficiency by integrating both internal and external knowledge, allowing the system to iteratively generate and refine planning outcomes\cite{yao2022react}. These efforts collectively highlight the growing influence of LLMs in various aspects of multi-agent systems, planning, and control, offering new possibilities for future advancements.

Another emerging area of interest is the integration of Large Language Models (LLMs) with graph-based tasks. As noted by Kambhampati \cite{kambhampati2024llms}, while the capabilities of LLMs might be somewhat overestimated in certain contexts, they can still provide significant assistance under specific conditions. Ge et al. highlighted key challenges that expert LLMs encounter, including issues related to scalability, nonlinear task planning, and the difficulty of quantitative evaluation \cite{ge2024openagi}. Some works demonstrated the potential of LLMs in solving a wide range of graph-related problems, from determining the connectivity between two nodes to enhancing the performance of Graph Neural Networks (GNNs) \cite{zhang2024can, wang2024can}. Experimental results consistently underscore the promise of leveraging LLMs’ inference capabilities in tackling graph-based challenges.


In this work, we incorporate LLMs into our ego-vehicle and service robot cooperation system to explore the improvements they can facilitate. Our approach involves graphs that represent the inferences of agents that can combine or split. Through effective communication and inference powered by LLMs, the system is designed to ensure that tasks proceed smoothly throughout the operation.

\section{Method}

\begin{figure*}[!t]
    \centering
    \includegraphics[width=1.0\textwidth]{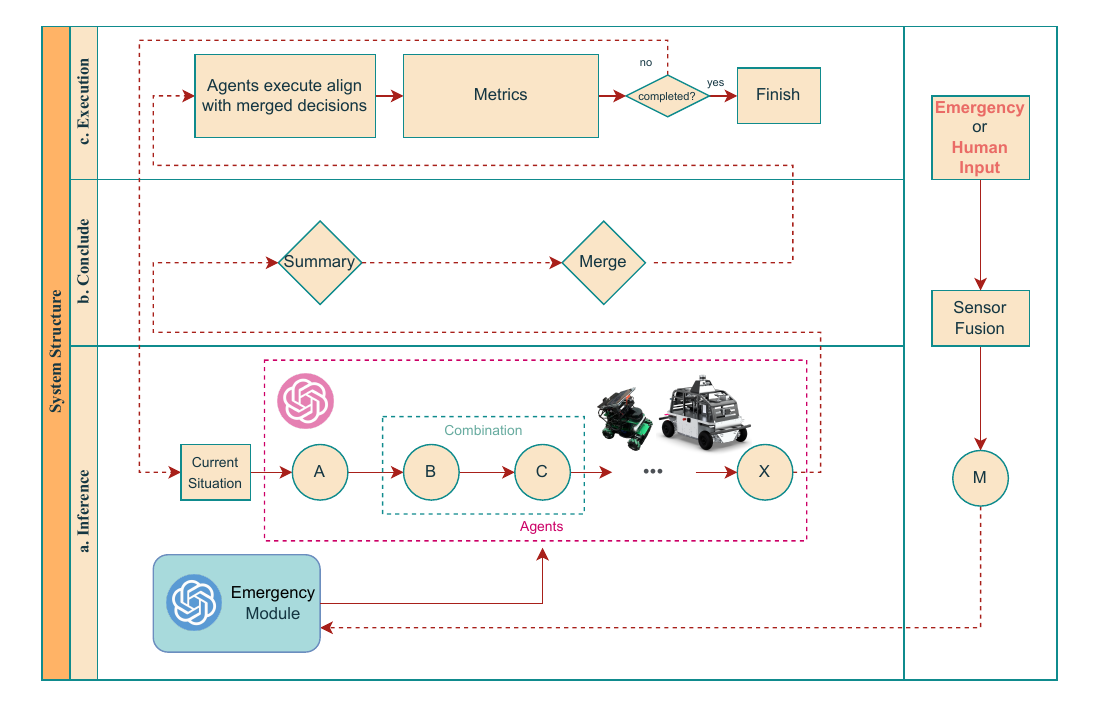}
    \caption{The structure of the CGoT framework. The system operates in a continuous loop between the inference, conclude, and execution phases until all tasks are completed. When an emergency event or human input occurs, the emergency handling module joins in the loop to ensure that task execution proceeds smoothly and efficiently.}
    \label{CGOT-structure}
\end{figure*}

\subsection{Problem Formulation}

Embodied agents can enhance their inference capabilities by organizing their knowledge and relationships in the form of a graph, as demonstrated in recent studies \cite{besta2024graph}. In this work, we consider a multi-agent system operating within a decentralized communication framework which we call Composable Graphs of Thoughts (CGoT), where agents utilize a Graph of Thoughts (GoT) as their primary form of inference. This system comprises ego-vehicles and robots with distinct functionalities, where agents are capable of dynamically forming new entities based on their contextual requirements. For instance, ego-vehicles can transport different robots to expedite task completion at specific locations. Additionally, the system is designed to operate effectively in dynamic environments, managing both certain tasks and uncertainties.

Building upon the concept of the Graph of Thoughts (GoT), the inference process of a multi-agent system performing tasks in a dynamic environment can be formalized as a tuple: (\emph{\textbf{G}}, \emph{\textbf{T}}, \emph{\textbf{L}}\textsubscript{$\theta$}, \emph{\textbf{I}}, \emph{\textbf{D}}). Here, \emph{\textbf{G}} denotes the collection of inference graphs for all agents, expressed as \emph{\textbf{G}} = \{\emph{\textbf{G}}\textsubscript{1}, \emph{\textbf{G}}\textsubscript{2}, ..., \emph{\textbf{G}}\textsubscript{n}\}, where each graph \emph{\textbf{G}}\textsubscript{i} is defined as a pair ({\emph{\textbf{V}}\textsubscript{i}}, {\emph{\textbf{E}}\textsubscript{i}}), representing the set of vertices and edges. To enhance system adaptability and scalability, as well as to streamline and refine the inference process, when certain nodes in the graphs represent the decision to combine agents, the combined agents are replaced by a new agent resulting from the combination.

\emph{\textbf{T}} represents the transformation of \emph{\textbf{G}}, denoted as \emph{\textbf{T}} = \{\{\{\emph{\textbf{G}}\textsubscript{n1}\}\textsuperscript{-}, \emph{\textbf{G}}\textsubscript{m1}\textsuperscript{+}\} ... \{\{\emph{\textbf{G}}\textsubscript{nk}\}\textsuperscript{+}, \emph{\textbf{G}}\textsubscript{mk}\textsuperscript{-}\}\}, where the agents {\emph{\textbf{G}}\textsubscript{n}} are combined into new agents {\emph{\textbf{G}}\textsubscript{m}} and {\emph{\textbf{G}}\textsubscript{m}} replaces {\emph{\textbf{G}}\textsubscript{n}} in the system’s inference, or the opposite. \emph{\textbf{L}}\textsubscript{$\theta$} represents the large language model (LLM) employed for inference, while \emph{\textbf{I}} refers to external inputs such as environmental disturbances or human interactions.

\emph{\textbf{D}} indicates the decisions made by agents after the inference process, and \emph{\textbf{E}} captures the state and progression of tasks. The entire process of task execution can thus be represented as follows:

The formulation in Equation \ref{G'} expresses the updated graph $\mathbf{G}^{' }$ as a function of the transformed graph $T(\mathbf{G})$, the environment state $\mathbf{E}$, and external inputs $\mathbf{I}$.  
\begin{align}\label{G'}
\mathbf{G}^{'} = \mathcal{F}\left(\mathbf{T}\left( \mathbf{G} \right), \mathbf{E}, \mathbf{I} \right)
\end{align}

Equation \ref{D=FG} defines the decisions made by the agents, represented as $\mathbf{D}$, which are determined by the output of the graph $\mathbf{G}$. This is expressed as 
\begin{align}\label{D=FG}
\mathbf{D} = Out\left( \mathbf{G} \right)
\end{align}
where $Out(\mathbf{G})$ determines the actions or choices the agents will take based on the inference graphs.

Finally, the environment state is updated
\begin{align}\label{E'}
\mathbf{E}^{' } = \mathcal{H}(\mathbf{E}, \mathbf{D}, \mathbf{I})
\end{align}
where $\mathbf{E}^{' }$ is defined as a function of the current environment state $\mathbf{E}$, the agents' decisions $\mathbf{D}$, and external inputs $\mathbf{I}$.

\subsection{System Structure}
Compared to traditional rule-based methods, one of the key advantages of embodied agents lies in their generalization capability and scalability. Rule-based systems often struggle with flexibility and adaptability, particularly when confronted with new tasks or dynamic environments. In contrast, embodied agents can adapt more effectively, making them ideal for complex, real-time applications. This work introduces a multi-agent embodied framework designed to handle specific system configurations, where agents of various types have the ability to physically combine. A prime example of such a system is an ego-vehicle and service robot collaboration, where ego-vehicles can transport robots. In this setup, ego-vehicles can choose to carry robots as needed, enabling more efficient task planning and accelerating the overall task execution.

The framework, depicted in Fig. \ref{CGOT-structure}, is structured into three layers: Inference, Conclude, and Execution. Each time the system state is updated or new inputs are received from the human operator or the environment, the agents begin the inference process. During this phase, they communicate and make decisions based on the current situation and the decisions made by other agents. Once inference is complete, the decisions are passed to the Conclude layer, where they are synthesized into a unified, executable result. Finally, in the Execution layer, the agents carry out the actions based on the merged decisions, continuing until the task is fully completed.

One of the distinguishing features of the proposed framework is the dynamic combination of agents within the system. When a physical relationship is established between two or more agents during task execution, their inference processes are also merged. This means that, during inference, these agents will function as a single entity until the relationship is dissolved. For instance, in the ego-vehicle-robot collaboration system described earlier, when an ego-vehicle carries a robot, the embodied agents of both the vehicle and the robot are combined into one. The newly formed agent inherits the capabilities of both original agents, enabling it to perform tasks that require the functionality of both.

In terms of embodied agent inference, the Graph of Thoughts (GoT) method plays a state-of-the-art (SOTA) role in this domain \cite{besta2024graph}. In the proposed framework, agents perform inference through a graph-based structure. Specifically, all known conditions are represented as initial nodes in the graph. Each agent starts its inference process from these nodes. As the inference progresses, intermediate nodes are generated, reflecting the evolving state of the task, until the final output is produced.

Whenever an output node in an inference graph is combined with another agent, the two agents merge into a single entity for the remainder of the process. The newly formed agent retains the functions and capabilities of all the combined agents. If, at any point, the inference graph of the new agent generates an output node indicating a discombination, the agent will split back into the two original agents, which will then resume their respective inference processes independently.

\section{Experiment}

\subsection{Basic Settings}

To evaluate the performance of the CGoT framework in solving multiple tasks within a dynamic environment, the experiments are conducted in a simulated park area comprising three buildings and a package delivery site. The tasks within this environment are divided into two categories: building cleaning and package delivery. The system consists of two ego-vehicles and two robots, each responsible for executing a series of tasks.

The two ego-vehicles are identical in capabilities; they are both capable of carrying robots and packages. One of the robots, referred to as Robot A, is responsible for assisting with package delivery into buildings, while the other, Robot B, is tasked with cleaning the buildings.

The cleaning task is simple: Robot B enters a building that has not yet been cleaned and performs the task within a single turn. The delivery task is more complex and consists of two stages: the first stage involves the ego-vehicle transporting the package from the package site to the designated building, while the second stage requires Robot A to move the package from the building's entrance into the building itself. Only when both stages are completed is the delivery considered finished for that building.

This experimental setup introduces a certain level of complexity and necessitates the collaboration of all agents within the system, making it an ideal scenario for testing the feasibility and efficiency of various planning methods. 

\subsection{Execution Efficiency}

In the experiments, several tests are designed to evaluate the feasibility of the CGoT framework. The environment consists of three buildings, each potentially assigned different tasks. 
These tests focus on evaluating the planning abilities of both frameworks. Under the same experimental settings, we examine the performance of both methods. The CGoT framework demonstrates comparable planning performance to the GoT framework without the inference-combination mechanism, as both frameworks exhibit strong generalization capabilities and adaptability in handling tasks.

\subsection{Token Consumption}
\begin{figure}[t]
    \centering
    \includegraphics[width=1.0\linewidth]{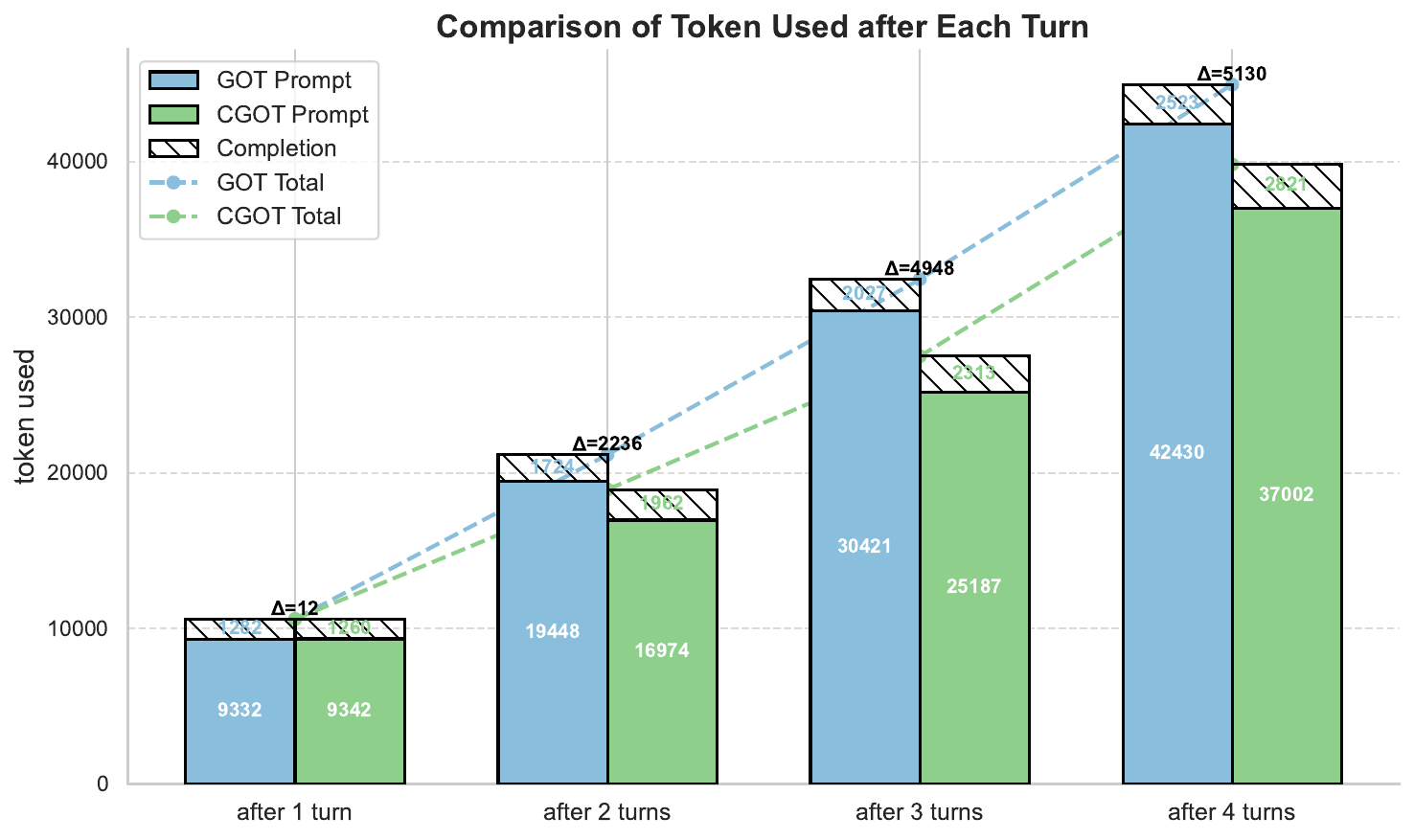}
    \caption{The contrast of GoT and CGoT on token consumption during a test. CGoT saves more tokens than GoT as the proceeding of the tasks execution.}
    \label{token2}
\end{figure}

However, the CGoT framework offers a notable advantage over GoT when it comes to token consumption. In traditional inference frameworks, agents perform inference independently, which leads to significant resource inefficiencies. In systems where agents have physical or signal-based interactions, such as the one proposed in this work, CGoT can significantly reduce token consumption while maintaining high standards of inference and communication. This advantage is evident in Fig.~\ref{token2}, where token consumption decreases progressively with each turn, especially as the number of agent combinations increases. As the system runs longer and more agents are combined, CGoT saves an increasing amount of tokens compared to the GoT framework.


\section{Conclusion}

This paper introduces the CGoT framework for multi-agent task planning in dynamic environments, combining the Graph of Thoughts (GoT) approach with dynamic agent interactions. The results demonstrate that CGoT achieves comparable planning performance to traditional methods and GoT without combination, while significantly reducing token consumption as agents collaborate.

The key advantage of CGoT is its adaptability to dynamic environments, assisting agents to combine and split based on situational needs. This flexibility, along with the reduction in communication overhead, makes CGoT more efficient for large-scale applications.

Future work could explore expanding CGoT to handle more complex environments and integrating advanced learning algorithms for continuous optimization. The framework shows strong potential for real-world applications in autonomous and collaborative systems.

\bibliographystyle{ieeetr}
\bibliography{mybib.bib}

\end{document}